\documentclass{elsart}

\usepackage{graphicx}
\usepackage{supertabular}
\usepackage{rotating}
\usepackage{slashed}

\usepackage{amssymb}
\usepackage{hyperref}

\def\aitalc{{\sc aITALC}}
\def\unix{{\sc Unix}}
\def\linux{{\sc Linux}}
\def\solaris{{\sc Solaris}}
\def\diana{{\sc Diana}}
\def\tedi{{\sc Tedi}}
\def\form{{\sc Form}}
\def\fortran{{\sc Fortran}}
\def\c{{\sc C}}

\def\make{{\sc Make}}

\def\topfit{{\sc Topfit}}
\def\qgraf{{\sc Qgraf}}
\def\LT{{\sc LoopTools}}

\def\ff{{\sc FF}}

\def\textmu{{$\mu$}}
\def \gf{\gamma_5}
\def \gi{\mathbf{1}}

\newcommand{\fig}[1]{Fig.~\ref{#1}}
\newcommand{\tab}[1]{Tab.~\ref{#1}}

\newcommand{\arxiv}[1]{\href{http://arxiv.org/abs/#1}{#1}}

\def\be#1\ee{\begin{equation}#1\end{equation}}
\def\bea#1\eea{\begin{eqnarray}#1\end{eqnarray}}
\def\bal#1\eal{\begin{align}#1\end{align}}
\def\btab#1\etab{\begin{table}[h]\begin{center}#1\end{center}\end{table}}
\def\bfig#1\efig{\begin{figure}[h]\begin{center}#1\end{center}\end{figure}}
\def\bi#1\ei{\begin{itemize}#1\end{itemize}}

\begin{document}
{\tt
\noindent DESY 04-226\\
\noindent SFB/CPP-04-63\\
\noindent hep-ph/0412047\\
\noindent December 2004\\
}
\begin{frontmatter}


 \title{An Integrated Tool for Loop Calculations: aITALC\thanksref{label1}}
 \thanks[label1]{This work was partially supported by the European's
   5-th Framework under contract HPRN--CT--2000--00149 Physics at
   Colliders and by the Deutsche Forschungsgemeinschaft under contract
   SFB/TR 9--03.} 
 \author{Alejandro Lorca and Tord Riemann}
 \ead{alejandro.lorca@desy.de,tord.riemann@desy.de}
 \address{Deutches Elektronen-Synchroton, DESY, Platanenallee 6, 15738
   Zeuthen, Germany}

\begin{abstract}
aITALC, a new tool for automating loop calculations in high energy
physics, is described. The package creates Fortran code for
two-fermion scattering processes automatically, starting from the generation and analysis of the Feynman graphs.
We describe the modules of the tool, the intercommunication between
them and illustrate its use with three examples.
\end{abstract}

\end{frontmatter}

\section{Program summary}
\bi
\item {\em Title of the program}: \aitalc\ version 1.0 (29 October 2004)
\item {\em  Catalogue identifier}:
\item {\em  Program obtainable from}: \url{http://www-zeuthen.desy.de/theory/aitalc}
\item {\em  Computer}: PC i686
\item {\em  Operating system}: GNU/\linux{}, tested on different distributions
  SuSE 8.2 and 9.1, Red Hat 7.2, Debian 3.0. Also on \solaris
\item {\em  Programming language used}: GNU \make, \diana, \form, \fortran77
\item {\em  Memory required to execute with typical data}: Up to about
  10 MB
\item {\em  No. of processors used}: 1
\item {\em  No. of bytes in distributed program, including test data, etc.}:
  369516 bytes, including tutorial in postscript format
\item {\em  Distribution format}: tar gzip file
\item {\em  High-speed storage required}: from 1.5 to 30 MB, depending on
  modules present and unfolding of examples
\item {\em  No. of cards in combined program and test deck}:
\item {\em  Keywords}: Radiative corrections, automatic calculation, Feynman
  diagram, computer algebra, \aitalc, \diana
\item {\em  Nature of the physical problem}: Calculation of
  differential cross sections for $e^+ e^-$ annhilation in one-loop approximation.
\item {\em  Method of solution}: Generation and perturbative analysis of
  Feynman diagrams with later evaluation of matrix elements and form
  factors.
\item {\em  Restriction of the complexity of the problem}: The limit
  of application is, for the moment, the $2 \to 2$ particles reactions
  in the electroweak standard model.
\item {\em  Typical running time}: Few minutes, being highly depending on the complexity of the process and the \fortran\ compiler.
\ei
\section{Introduction}
\aitalc\ is designed as a tool to perform automated perturbative
higher order calculations of cross sections in high energy physics.

The main goal is to create numerical programs directly from Feynman
rules. It is actually developed for few known models like the
electro-weak standard model (EWSM) and quantum electro-dynamic (QED),
and the limit of application is, for the moment, the $2 \to 2$
particles reactions involving only external fermions including the
one-loop virtual corrections and the soft photon bremsstrahlung.

It is expected from the user to cook his or her own process within a set of basic and/or advanced ingredients. As a reward to such effort comes the output of \aitalc: Fast and reliable \fortran\ code with differential and integrated cross sections, analytical expressions for the transition amplitude and nice drawings of the individual Feynman graphs.

The tool was intended to integrate only free-of-charge packages existing on the market. As the calculation proceeds, in a modular fashion, it profits from the \diana\ package \cite{Tentyukov:1999is} (based itself on the \qgraf\ \cite{qgraf} code) for the generation and analysis of Feynman graphs, the \form\ \cite{Vermaseren:2000nd} language when dealing analytically with the large expressions in the amplitudes and, finally, the \LT\ \cite{Hahn:1998yk} library (also integrating the \ff\ package \cite{vanOldenborgh:1990yc}) for the numerical calculation of the loop integrals.

This package is developed from the evolution of the static \fortran\ code \topfit\ \cite{Fleischer:2003kk}, intended for the calculation of top-pair production. Some of the features then available such as the hard-photon bremsstrahlung or separate weak-QED contributions were discarded in pro of a full automation, higher technical precision and flexibility for many other processes.
\section{Technical requirements and configuration}
We present a list of the necessary conditions to install and run \aitalc{} on a computer.
\begin{itemize}
\item[]{\bf System compulsory requirements:}
\begin{itemize}
\item A computer under \linux%
\footnote{Installation on other \unix{} systems might be possible if there are already some modules present. The package has been successfully build in {\sc Solaris} for testing purposes.}
  operating system with standard%
\footnote{Free packages can be obtained from \url{http://www.gnu.org} under the names {\tt make} and {\tt gcc}.}
\begin{itemize}
\item {\sc GNU }\make{} utility
\item \c{} compiler
\item \fortran 77 compiler (with preprocessing ability)
\end{itemize}
\end{itemize}
\item[]{\bf Suggested optional features:}
\begin{itemize}
\item CPU speed $\ge$ 333 MHz
\item RAM Memory $\ge$ 128 MB
\item At least from 1.5 to 10 MB%
\footnote{The fewer modules you have already installed the larger the size for including them. Another extra 30 MB will be needed when building the three examples proposed.}%
 free space on disk for installation 
\item Further limitations given by the individual modules
\end{itemize}
\end{itemize}

The package is licensed under the GNU general public license and can
be found under the official web page
\url{http://www-zeuthen.desy.de/theory/aitalc}. The documentation
\cite{aitalctutorial} can be accessed within the package distribution
in postscript format or at the web page also as {\tt pdf}.

The installation includes an accommodation to your system settings via
{\tt con}\-{\tt fig}\-{\tt ure} and the later extraction of individual packages (if
needed) and examples. This process is also controlled by a {\tt make}
instruction, conforming to the standards of the \unix\ packages.
\section{Makefile like an environmental language}
Automation requires effective inter--communication and a smart way to organize the tasks. This second concept has been widely exploited through the Makefile environment, being its conceptual organization a pro in order to build different sections of the calculation and provide a chain dependency between files and modules.

The Makefile environment accomplishes
\bi
\item The simplification of user interface
\item Building the sections {\tt tree}, {\tt loop} and {\tt fortran}
\item Running the driver process file
\item Writing the intermediate information \diana{} $\rightarrow$ \form{}
  $\rightarrow$ \fortran{} in a row
\item Compiling the numerical program leading to the final results
\ei
in a modular fashion. The execution is carried out sequentially following the control flow structure designed in \fig{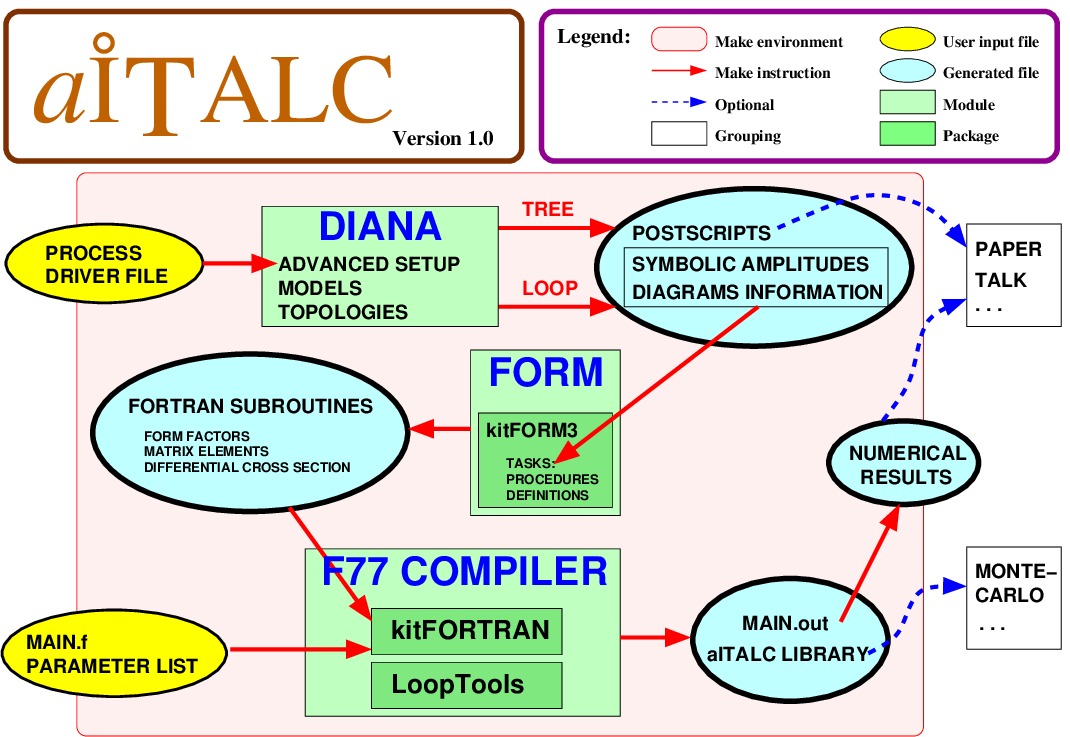}.

\begin{figure}[t]
\begin{center}
\includegraphics[scale=1]{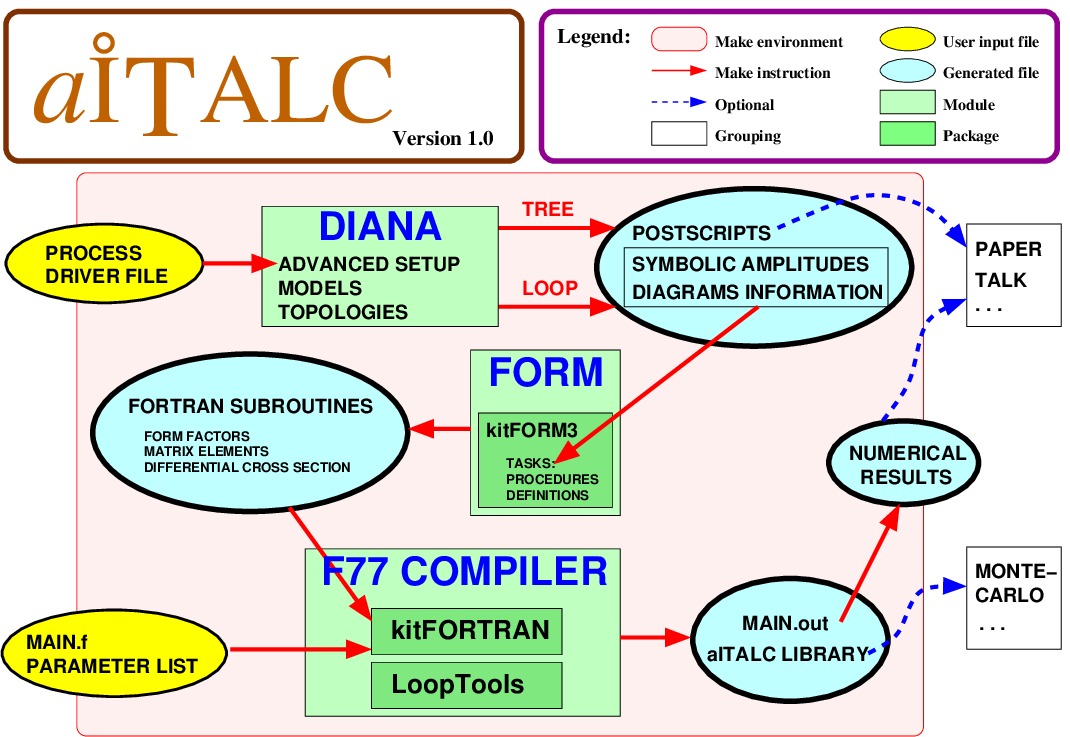}
\caption{Logical flow chart structure for \protect\aitalc.}
\label{aITALC_flowchart-v1.0.eps}
\end{center}
\end{figure}

Despite of the simplicity, the idea of calculating a process by a {\tt
  make} instruction has another major advantage: reproducibility,
being understood as the ability to recreate once and again similar
processes for testing or cross-checking purposes. On the contrary,
customizing of an specific process turns out to be a task spread among different parts: driver file, add-on files for neglecting the default ones and specific commands in the main fortran programs.

We will dedicate next parts to a deeper discussion of how each module addresses its tasks and the description of the {\it modus operandi}.
\section{Feynman diagram generation with \protect \diana}\label{sec:diana}
Let us introduce some basic concepts about \diana. If we are thinking of some process, a very intuitive approach is to draw some Feynman diagrams to get a feeling and support the calculation which comes later.

Such a task is very likely to be implemented with the help of computers since it is basically a combinatorial problem. Once we define a model, the external particles, and the level of loops, the problem can be solved. Indeed, \diana{} will do it for us.
\subsection{Basic driver file: Input}
The driver file (e.g. {\tt process.ini}) serves as basic input. It contains primarily:
\bi
\item physical model,
\item incoming and outgoing particles,
\item process control flags and options.
\ei

These items are used in the \diana{} ``create'' file, that will serve as main user input stream for \diana\ at each level, i.e, {\tt tree} or {\tt loop}. Only the number of loops is left to be arranged and written by the {\tt Makefile} on the ``create''.
For example, it is compulsory to check and agree with the five lines
\begin{verbatim}
SET _processname=muon_production
...
\Begin(model,EWSM.model)
...
ingoing le(;p1),Le(;p4);
outgoing lm(;-p2),Lm(;-p3);
...
options=notadp,onshell;
\end{verbatim}
in {\tt process.ini} before creating any process. For the case of
\textmu-pair production, as considered in the previous lines, the full
driver file is shown in the \fig{process.eps}. Only one remark at this
point: special care is needed with the momenta definitions. They all
should be assigned to be incoming and clockwise indexed (e.g. {\tt p1}
addressed to be the first incoming particle, {\tt -p2} to be the first
outgoing one and so on), so we ensure a matching with the
convention used in the predefined topologies.

\bfig
\includegraphics[scale=0.48,angle=270]{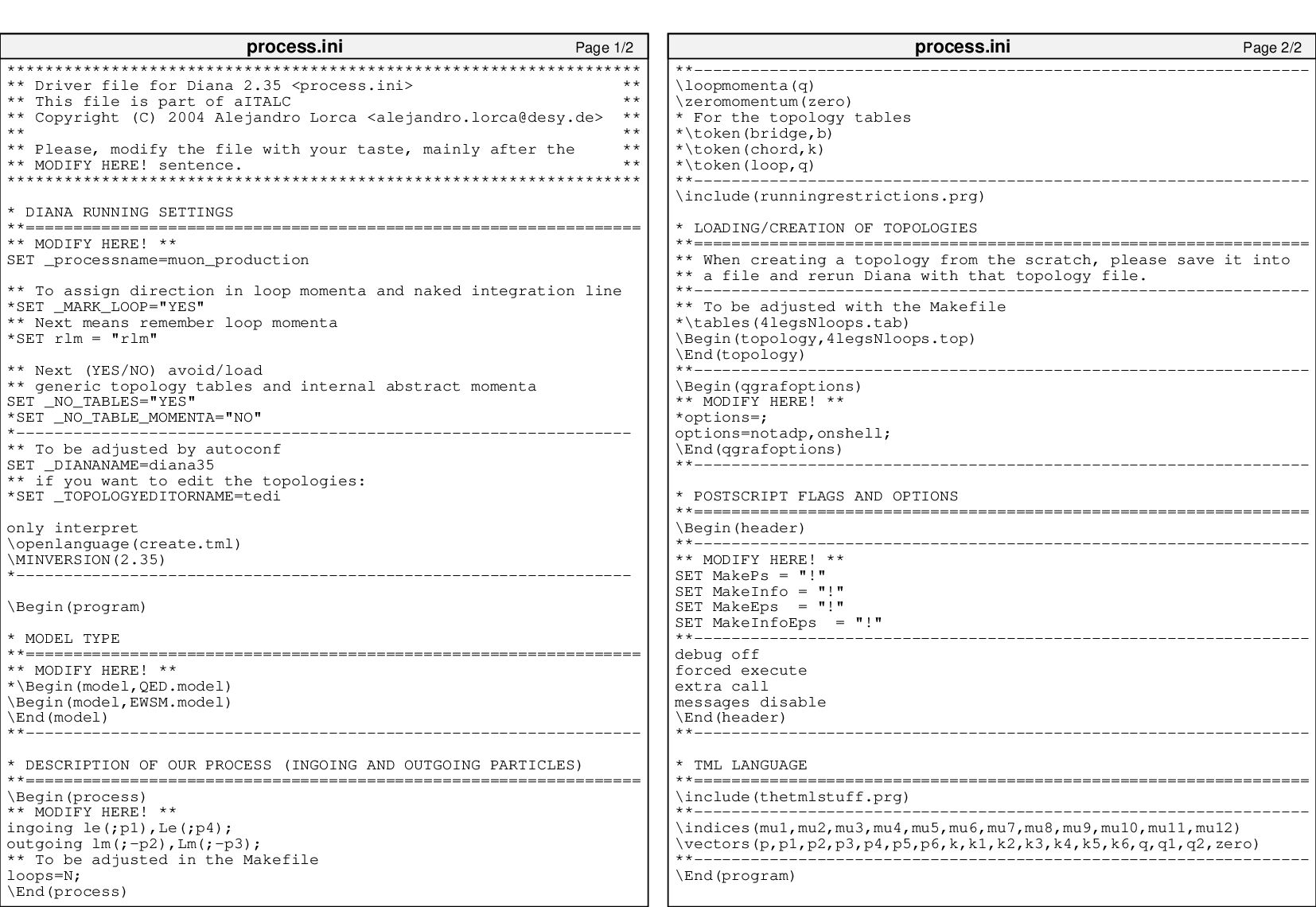}
\caption{Snapshot for {\tt process.ini} driver file.}
\label{process.eps}
\efig

The comment character is {\tt *} and the
{\tt ** MODIFY HERE! **} sentences indicate where some modifications
are accepted. Regarding the flags for the initial settings, the
topology editor (\tedi{}) can be launched at running time by
uncommenting the line:\\ {\tt *SET} {\tt \_TOPOLOGYEDITORNAME} {\tt="tedi"}.
\subsection{Implemented models}
Once the driver file is fully set up, \diana{} is able to offer some
 of the information to \qgraf{} and, by looking to the allowed
 couplings given in the model file, create the possible diagrams
 appearing at each level. The presently available model files are {\tt
   QED.model} and {\tt EWSM.model}. They are summarized in \tab{qedtable} and \tab{ewsmtable}.

Both of these models share basic properties according to their phenomenological content. We can distinguish between the following sections:
\bi
\item Propagator definitions, with general structure\\
{\tt [} Particle {\tt ,} Antiparticle {\tt;} Prototype {\tt;} Function
{\tt(} Prop-label {\tt,} arguments {\tt)*} factor {\tt;} Mass-sq. {\tt ]}
\bi
\item Boson, e.g. 
{\tt [Wm,Wp ;W; VV(num,ind:1 ,ind:2 ,vec, 2)*i\_ ;MMW]}
\item Fermion, e.g.
{\tt [le,Le;l; FF(num,fnum,vec, Mle)*i\_; MMle]}
\item Ghost, e.g.
{\tt [ghZ,GhZ; Z ; SS(num,1)*i\_; MMZ]}
\ei
\item Vertex definitions, as\\
{\tt [}Interacting particles{\tt ;;}Function{\tt(}arguments{\tt)*}factor{\tt ]}
\ei
in which the specific Feynman rules will act later. They are, to some extent, part of the model but, from a computational point of view, they belong to the module controlled by \form{} programs and therefore encoded in a procedure ({\tt ctfeynmanrules}).
\subsection{Topology editor}
\tedi{} is designed to allow the user ultimate flexibility on topology
naming, shaping and even defining momenta. It was developed together
with \diana\ as a tool for dealing with topological issues of the
Feynman diagrams, and requires the X window server being running in
your system. The `windows' and `tab' structure renders an intuitive use, as it is shown in \fig{tedisnapshot}. Furthermore, help and printing facilities are also embedded in the program.

\bfig
\includegraphics[scale=0.75]{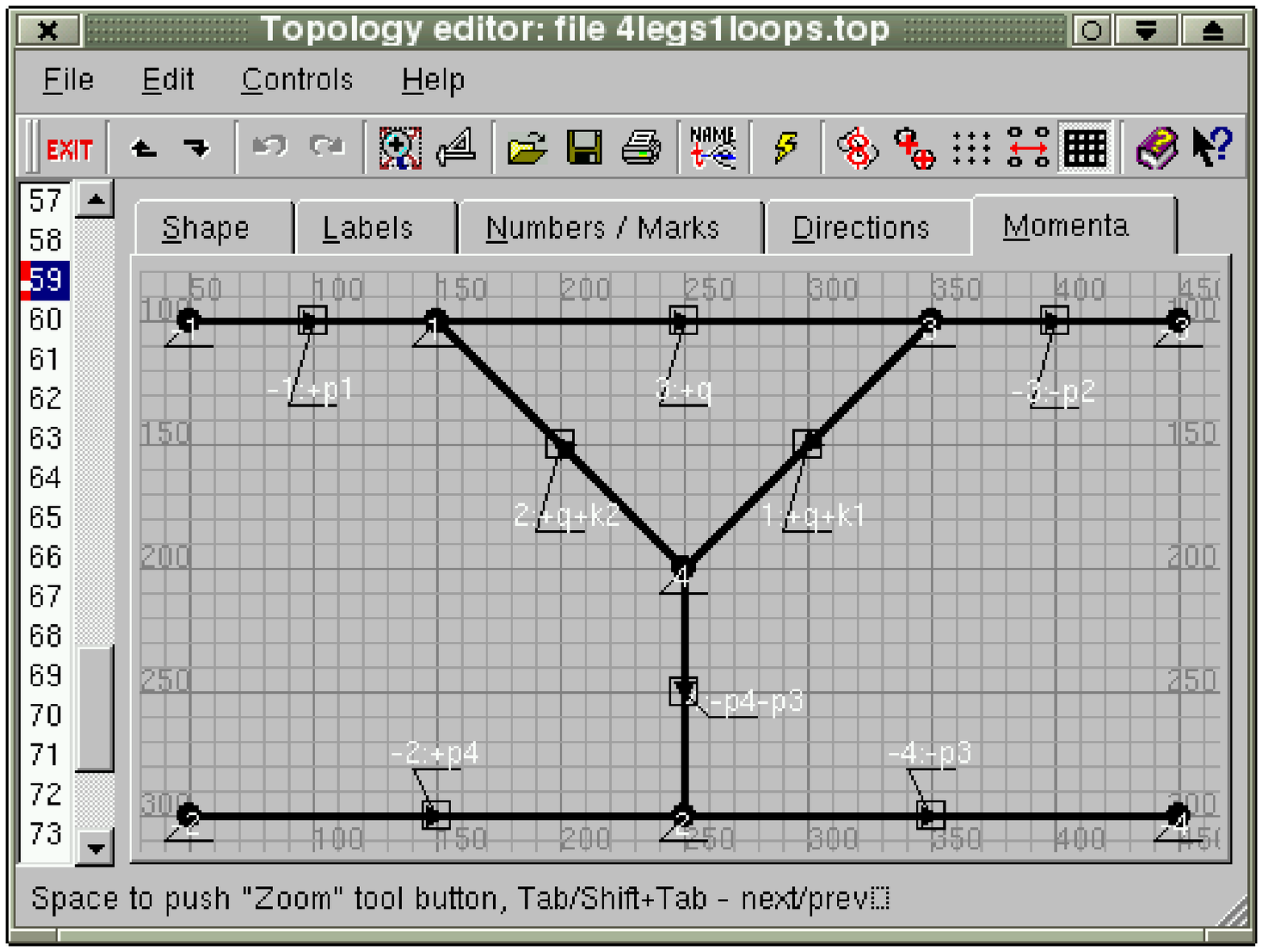}
\caption{Snapshot for the momenta tab in \protect\tedi{}.}
\label{tedisnapshot}
\efig

The content of the five tabs has to do with
\bi
\item Shape: handles with the aesthetics of the Feynman diagrams. Here the users can reshape the different legs and loops of the topology according to an optimized algorithm or their own taste. Aligning to a grid is also possible by adapting the alignment distance and toggling the key points between nodes for each object.
\item Labels: External legs, internal propagators and vertices are given a label in \diana. The location of this label might be varied with respect to the graph to avoid non-legible superpositions.
\item Numbers/Marks: Every line has an index that is negative for
  external legs and positive for internal propagators. Working with this tab one can ensure to give the highest index values to a loop or to fixed-momenta lines.
\item Directions: The sign of the momentum flow in each internal line can be flipped by clicking on the arrow with the right-button mouse.
\item Momenta: Alternative symbols can be arranged for the momenta attribute.
\ei

Apart from this main five features, a bar on the left scrolls between the different topologies that were generated for the specific process.

A deeper discussion of the possibilities of \tedi\ is out of the focus of a standard \aitalc\ description. Indeed it has been chosen that the default running of any of the examples does not require the topology editor at any moment. We will close this summary with the \fig{Bhabha_topologies} in order to give the user an impression of the standard graphical output of \tedi.

\bfig
\includegraphics[scale=0.35]{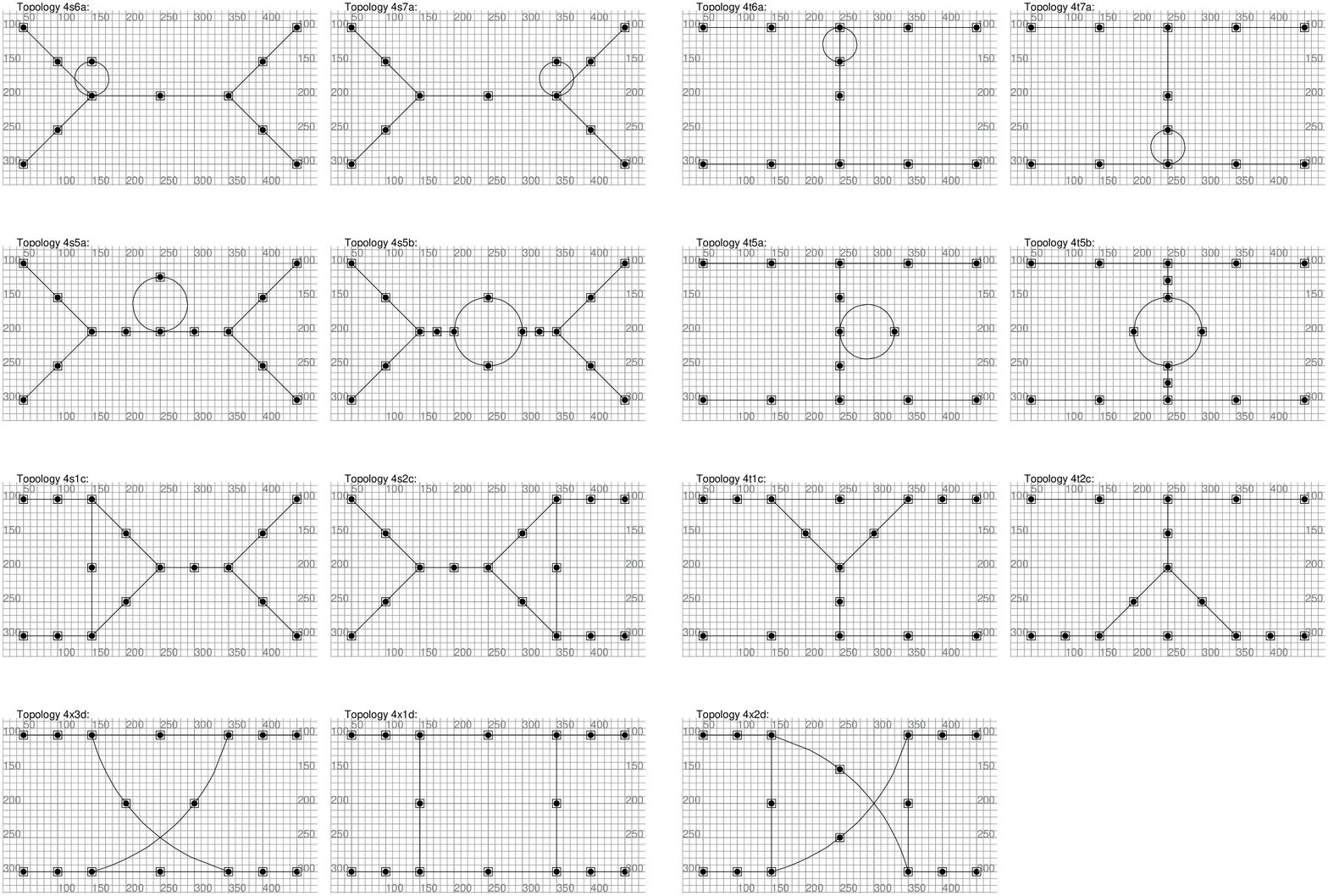}
\caption{Collection of topologies given by \protect\tedi{} for the Bhabha scattering at one-loop level. Tadpoles on vertices account for counterterms diagrams.}
\label{Bhabha_topologies}
\efig
\subsection{Advanced setup}
But what if we want to exclude some couplings or modify the aspect of the diagrams?

After the installation of \aitalc\, it is also possible to use
\diana\footnote{This extends also to \form{} and \LT{}.} as an
independent package. We will show shortly the tested modifications that can be performed at this level.

After running an example (e.g.\ {\tt muon\_production}), inside the {\tt tree} directory (also the {\tt loop} directory in other examples) it is possible to select some files to be considered instead of the default ones coming from the following directory: {\tt \$AITALCHOME/diana/prg/}. This is automatically done by placing those files on the {\tt examples/muon\_production/} directory and remaking the process. When the process has to be run for the second time, the instructions are: {\tt make clean} and then {\tt make} again.

We do not suggest to copy and modify any other file from {\tt prg}
since they are mostly functions using advanced \diana{} declarations. The following files are welcomed to be modified:
\bi
\item {\tt particleaspect.prg}: Defines how every propagator and
  particle label looks like in the {\tt .ps} and {\tt .eps}
  files. Syntax is not described (just a couple of comments), but
  intuitive. The field label {\tt CT} is assigned to counterterms.
\item {\tt runningrestrictions.prg}: Fields or couplings can be
  excluded here, by uncommenting or replacing them where desired%
\footnote{Warning: As soon as you may discard some fields or couplings, be aware that the renormalization of parameters and fields was fixed and completed within each model and does not know about your intentions, so performing a consistency check of the results is mandatory.}%
.
\ei
\subsection{What do we get?}
\begin{table}[th]
\caption{Listing for the {\tt tree} directory inside the muon  production example.}
\label{muontree}
\begin{center}
\begin{tabular}{|c|}
\hline
\begin{minipage}[l]{35em}
{\small
\begin{verbatim}

alorca@linux:~AITALCHOME/examples/muon_production/tree> ls -XFl
total 183

drwxr-xr-x 2 alorca th   2048 Oct 14 15:02 EPS/
drwxr-xr-x 2 alorca th   2048 Oct 14 15:02 FFmuon_production/
drwxr-xr-x 2 alorca th   2048 Oct 14 15:02 InfoEPS/
-rw-r--r-- 1 alorca th   3291 Oct 14 15:02 Makefile
-rw-r--r-- 1 alorca th   3710 Oct 14 15:02 muon_production.cnf
-rw-r--r-- 1 alorca th     20 Oct 14 15:02 fermioncurrentsnames.in
-rw-r--r-- 1 alorca th   3870 Oct 14 15:02 muon_production.in
-rw-r--r-- 1 alorca th   4531 Oct 14 15:02 Make_kit.log
-rw-r--r-- 1 alorca th    149 Oct 14 15:02 Makefile.log
-rw-r--r-- 1 alorca th   1088 Oct 14 15:02 do_amplitude.log
-rw-r--r-- 1 alorca th    245 Oct 14 15:02 joinff.1.log
-rw-r--r-- 1 alorca th   1059 Oct 14 15:02 joinff.log
-rw-r--r-- 1 alorca th    629 Oct 14 15:02 joinkinematics.log
-rw-r--r-- 1 alorca th 121102 Oct 14 15:02 joinmm.log
-rw-r--r-- 1 alorca th  14779 Oct 14 15:02 muon_production.ps
-rw-r--r-- 1 alorca th  18277 Oct 14 15:02 muon_productionInfo.ps

\end{verbatim}
}
\end{minipage}\\
\hline
\end{tabular}
\end{center}
\end{table}

Let's have a look into the level directory (e.g.\ {\tt tree/} or {\tt
  loop/} as in \tab{muontree}): the organizer file is called {\tt Makefile}. It has the instructions to create, organize and fill the different directories by given commands under the \make{} environment. This means that in order to build the process, only a {\tt make} instruction is needed. Depending on the process, modifications, hardware and compiler the time for each module varies from seconds till several minutes%
\footnote{If large files are to be compiled, occasionally the system might overload. Everything taking longer than half an hour is suspicious of going awry.}%
. At the beginning and end of each module, time stamps are placed in the {\tt Makefile.log} file at \tab{muontree}.

\begin{figure}[h]
\begin{tabular}{ll}
a) & b)\\
\raisebox{1pt}{\includegraphics[scale=0.275]{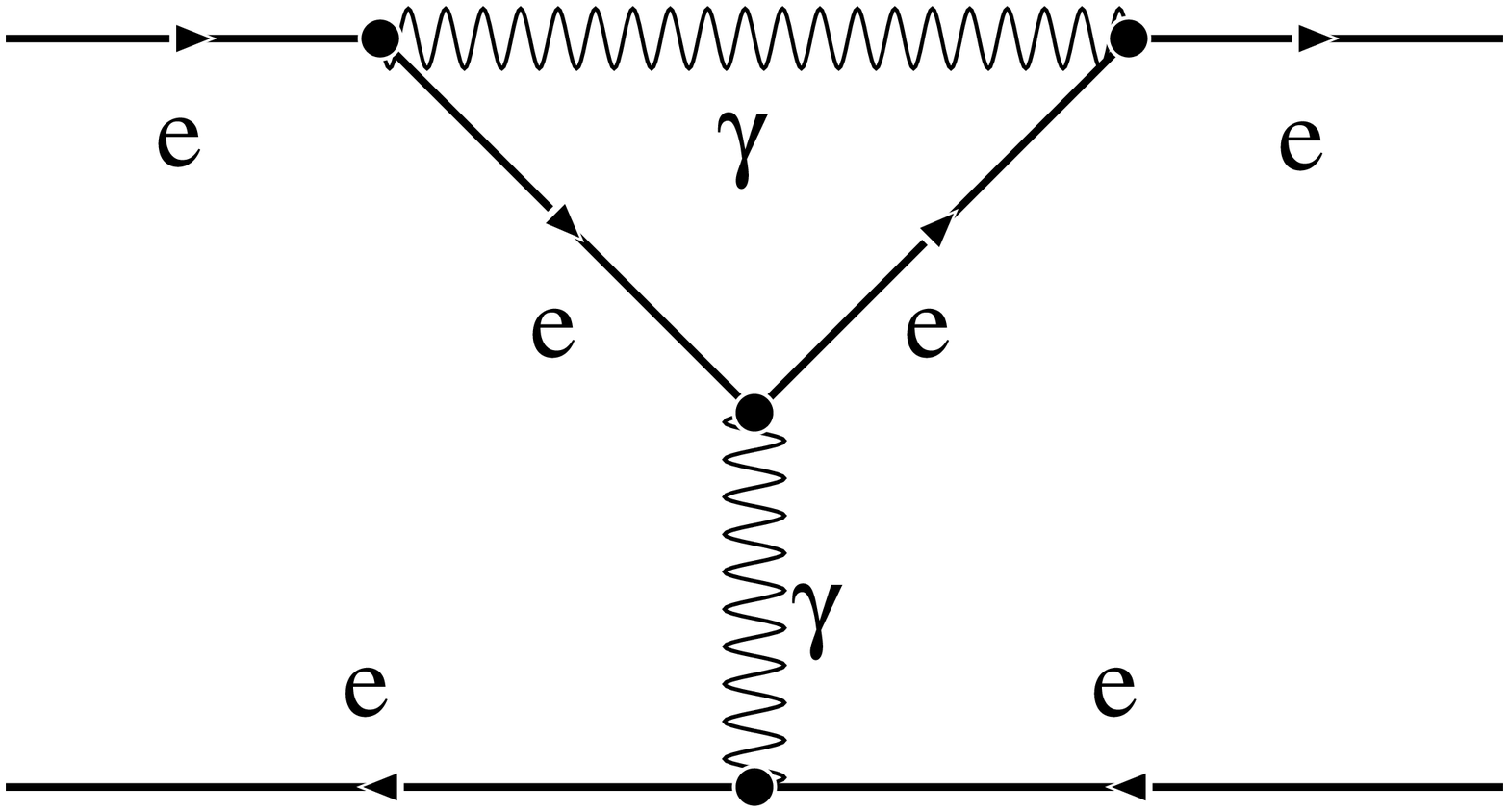}}&
\includegraphics[scale=0.26]{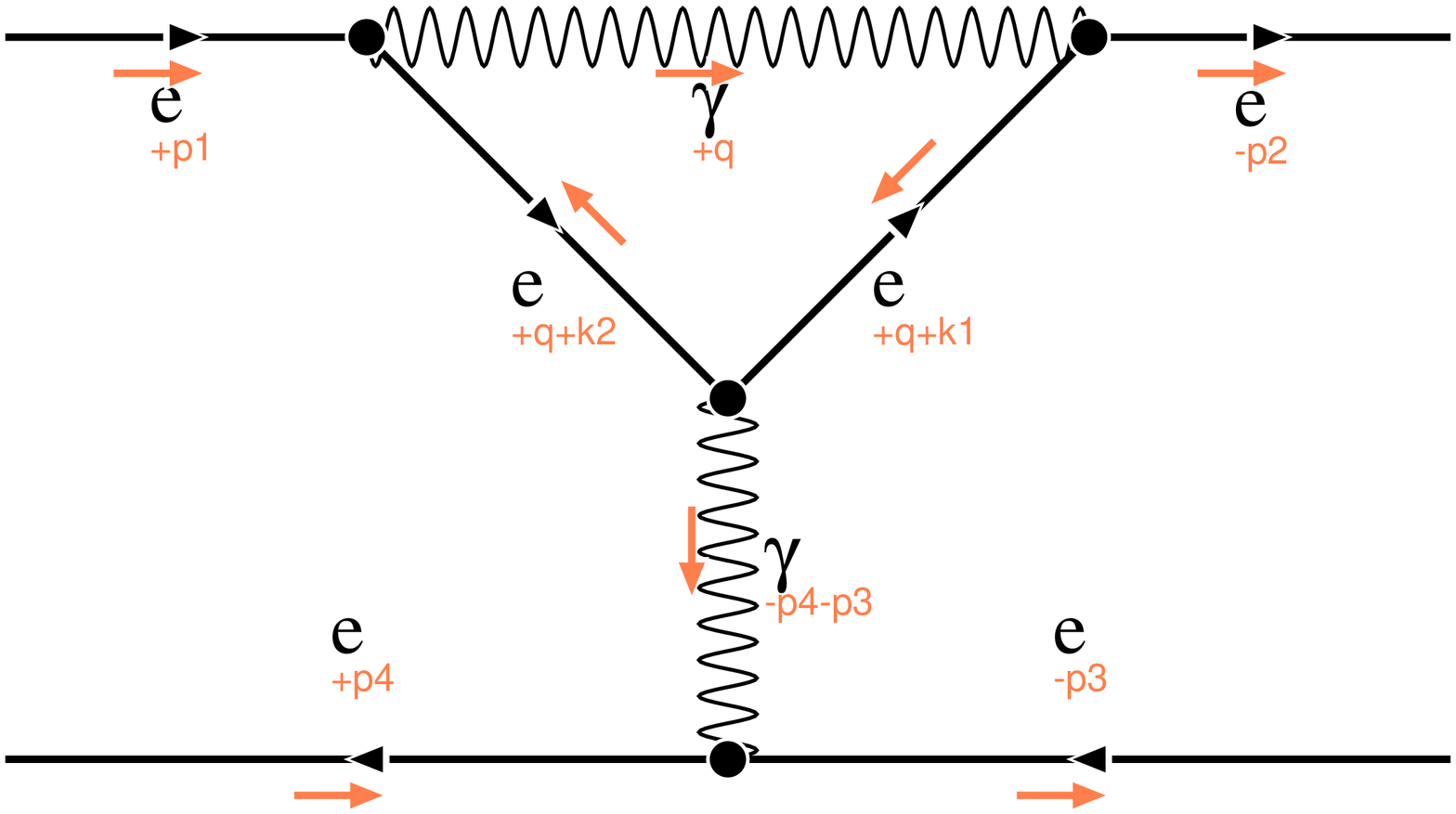}
\end{tabular}
\caption{Encapsulated postscripts diagrams for a typical diagram in Bhabha scattering. a) Simple diagram from {\tt EPS/} directory, b) also with momenta distribution from {\tt InfoEPS/}.}
\label{epsdiagrams}
\end{figure}

The directory also offers \form{} processed files that we will discuss later. At the moment we can concentrate in just two aspects:
\bi
\item {\bf Graphical representation} \\
The global files {\tt \$processname.ps} and {\tt \$processnameInfo.ps}
summarize all the Feynman graphs appearing at the process. In the {\tt
  EPS/} (and {\tt EPS}\-{\tt Info/}) directories one may find {\tt .eps} files with all individual diagrams (and mo\-men\-ta-label definition). \fig{epsdiagrams} represents the two types of output for a single vertex diagram for the $t$-channel exchange in Bhabha scattering.
\item {\bf Diagram info}\\
As it comes out of \diana, the file {\tt \$processname.in} contains the required information about both, global process and individual diagrams. Besides a compulsory expression for the amplitude, specially useful for the \form{} routines are the preprocessed variables defined for each diagram. Below we show the connection between the \fig{epsdiagrams} and the corresponding output in {\tt \$processname.in}. 
\ei

\begin{verbatim}
*--#[ n11:
****(qgraf number 11)
**** (diagram 11)

#define imm1 "MMle"
#define q1 "-q-k1"
#define loopline1 "1"
#define imm2 "MMle"
#define q2 "-q-k2"
#define loopline2 "1"
#define imm3 "MMA"
#define q3 "+q"
#define loopline3 "1"
#define imm4 "MMA"
#define q4 "+p4+p3"
#define loopline4 "0"

#define k1 "(+p2)"
#define k2 "(-p1)"
...
#define DIAGRAM "11"
#define COUNTER "11"
#define LINE "4"
#define FERMIONLOOP "0"
#define TOPOLOGY "4t1c"
#define CHANNEL "t"
#define SKELETON "1"
#define LOOPTYPE "c"
#define LOOPTYPENUMBER "3"
#define PROTOTYPE "llaa"
#define FERMIONCHANNEL "T"
#define l1 "1"
#define l2 "2"
#define l3 "1"
#define l4 "2"
*-----------------------------------------
g [Amplitude,11] =
       (-1)*F(3,1,mu2,1,0,1)*(-i_)*e*Qle*FF(1,1,-q-k1,Mle)*i_*
       F(4,1,mu4,1,0,1)*(-i_)*e*Qle*FF(2,1,-q-k2,Mle)*i_*
       F(1,1,mu1,1,0,1)*(-i_)*e*Qle*F(2,2,mu3,1,0,1)*(-i_)*e*Qle*
       VV(3,mu1,mu2,+q,0)*i_*VV(4,mu3,mu4,+p4+p3,0)*i_;

*--#] n11:
\end{verbatim}

\section{Algebra manipulation: Library in \protect\form}\label{sec:form}

The {\tt form} module acts as a bridge connecting the world of the
purely symbolic representation of the Feynman diagrams, directly read
off the lagrangian, and the numerical description of a scattering problem of real particles.

When \diana\ has finished one of the levels, a complete description of
the process is obtained, but still in an encoded language. Different \form{} programs will prepare the amplitudes in order to render numerically evaluable expressions. Tracing, commuting $\gamma$ matrices, applying the Dirac equation to external fermions or introducing the Mandelstam variables is part of this module.

There are basically two kind of programs: The main routines and the procedures. We describe the four routines:
\bi
\item {\tt do\_amplitude.frm}: Attempts to simplify the individual amplitude for each diagram, extracting out the form factors in a given basis of matrix elements.
\item {\tt joinkinematics.frm}: Defines how the differential cross section should be composed in terms of fermion currents ({\tt S}, {\tt T} or {\tt U}), the normalization of the incoming flux in the cross section formula, soft photon emission and the definition of the Mandelstam variables.
\item {\tt joinmm.frm}: The standard set of matrix elements is considered by this program, giving, as output, the multiplication of any two elements appearing in the amplitude decomposition.
\item {\tt joinff.frm}: Once all the diagrams were considered, this part pastes together the form factor contributions sorted by topologies. The final translation into \fortran{} code is also done at this step. 
\ei

\subsection{Procedures in {\tt kitFORM3/}}

The listing of the procedures is long, and the description of each of them is shortly presented in \tab{procedurelist}.

At least we would like to draw the attention to one option regarding the neglect of the external masses of particles (e.g. light fermion masses). This can be achieved including the adapted {\tt Neglectedmasses.inc} file into the main directory, together with the driver file {\tt process.ini}.

Coming back to our previous listing of the directory at
\fig{muontree}, in the files {\tt do\_am}\-{\tt pli}\-{\tt tude.log} and {\tt
  joinmm.log} we find information about the individual evaluation of
each diagram and the kinematical factor corresponding to the cross
product of matrix elements, respectively. Also, extra functions
related to soft photon emission, flux factors and Mandelstam variables
can be found in {\tt joinkinema}\-{\tt tics}\-{\tt .log}. As an intermediate step, individual amplitudes are stored in the {\tt FF\$processaname/} directory.

\section{Getting numbers: the \protect\fortran\ code}\label{sec:fortran}

So far the expressions that were generated contain large amount of
analytical work on them. Now one has to execute them and evaluate the
parameters, loop integrals and finally retrieve finite formfactors
that, clustered by topologies, will allow together with the matrix
elements to calculate the cross section for a given process.

The structure of the {\tt fortran} module can be outlined as follows:
\bi
\item{\tt main.f, parameterlist.hf}: they both give the user access
  to modify the output structure of the program and the explicit
  values of the model respectively. In the {\tt main.f} file we find
  many settings as logical flags, input of kinematics and output control.
\item{\bf \sc kitFORTRAN}: consists of the full set of \fortran{}
  subroutines and functions needed to compile the {\tt main.f}
  program.
We can divide them into two categories, according to the role they
play during running time.
\bi
\item {\bf Global} are those that were fixed at installation time
and considered to be process independent. They are stored at {\tt \$AITALCHOME/fortran/src}
\item {\bf Local} are written at running time by the {\tt tree} and
{\tt loop} modules and depend explicitely on the application. They get
placed at {\tt \$processname/}\-{\tt fortran/}\-{\tt src}.
\ei
\item{\bf \LT}: this library is called with the purpose of numerically
  calculating the most of the loop integrals.
\ei

\subsection{Understanding and controlling the output}\label{subsec:output}

A sample output are the Tabs.~\ref{main-bhabha_QED-long.log} and
\ref{main-leLe.bS-long.log} with differential cross sections (given
always in pb). The first column corresponds to the beam energy
($\sqrt{s}=\mathtt{setsqrtsman}$) given in GeV, the second one to the
$\cos\theta$ ({\tt setcost}), $\theta$ being the angle between the
three-momenta $\mathbf{p_1}$ and $-\mathbf{p_2}$.
Later on we have successively the columns of the Born approximation
({\tt Born}), the interference terms of the tree and loop levels ({\tt
  loop}) and the soft photonic
corrections ({\tt soft}). These three columns (third, fourth and fifth)
are summed into the sixth column ({\tt B+corr}) with the finally corrected cross
section. The last column {\verb+loop^2+} contains the one-loop squared terms in the perturbative approach.
The maximum energy fraction that the soft photon may gain out of
$\sqrt{s}$ is limited by the variable {\tt setfracomega}.

\begin{table}[th]
\begin{minipage}{\textwidth}{
\scriptsize
\begin{center}
\caption{File {\tt main.log}, for the {\tt bhabha\_QED} example.}
\label{main-bhabha_QED-long.log}
\begin{verbatim}
#=============================================================================================
# aITALC Version 1.0 (29.10.2004) by A.Lorca -- T.Riemann   
#---------------------------------------------------------------------------------------------
#sqrtsman      cost      dcs(Born)     dcs(loop)     dcs(soft)     dcs(B+l+s)    dcs(loop^2)
 0.5000000E+03 -.9000000 0.5238736E+00 0.1045142E+00 -.2193133E+00 0.4090746E+00 0.9235446E-02
 0.5000000E+03 -.5000000 0.6116007E+00 0.1767866E+00 -.2922801E+00 0.4961073E+00 0.1727483E-01
 0.5000000E+03 0.0000000 0.1172536E+01 0.3974063E+00 -.6065847E+00 0.9633579E+00 0.4206711E-01
 0.5000000E+03 0.5000000 0.5504406E+01 0.2126643E+01 -.3064635E+01 0.4566414E+01 0.2405476E+00
 0.5000000E+03 0.9000000 0.1891184E+03 0.8703341E+02 -.1165000E+03 0.1596518E+03 0.1089656E+02
#=============================================================================================
#Energy given in GeV, Cross sections in pb, setfracomega= 0.1000000
\end{verbatim}
\end{center}
}
\end{minipage}
\end{table}

\begin{table}[t]
\begin{minipage}{\textwidth}{
\scriptsize
\begin{center}
\caption{File {\tt main.log}, for the {\tt leLe.bS} example.}
\label{main-leLe.bS-long.log}
\begin{verbatim}
#=============================================================================================
# aITALC Version 1.0 (29.10.2004) by A.Lorca -- T.Riemann
#---------------------------------------------------------------------------------------------
#sqrtsman      cost      dcs(Born)     dcs(loop)     dcs(soft)     dcs(B+l+s)    dcs(loop^2)
 0.5000000E+03 -.9000000 0.0000000E+00 0.0000000E+00 0.0000000E+00 0.0000000E+00 0.6534118E-08
 0.5000000E+03 -.5000000 0.0000000E+00 0.0000000E+00 0.0000000E+00 0.0000000E+00 0.7095421E-08
 0.5000000E+03 0.0000000 0.0000000E+00 0.0000000E+00 0.0000000E+00 0.0000000E+00 0.1048690E-07
 0.5000000E+03 0.5000000 0.0000000E+00 0.0000000E+00 0.0000000E+00 0.0000000E+00 0.2084942E-07
 0.5000000E+03 0.9000000 0.0000000E+00 0.0000000E+00 0.0000000E+00 0.0000000E+00 0.4643956E-06
#=============================================================================================
#Energy given in GeV, Cross sections in pb, setfracomega= 0.1000000
\end{verbatim}
\end{center}
}
\end{minipage}
\end{table}

If the calculation is correct, all the columns should be constant\footnote{Warning: as a result of numerical variation, the round off of intermediate results will modify the last digits in your results. Such variation increases at the collinear cases due to cancellations.}
under variation of the ultraviolet (UV) parameter (you can check this
by turning on the flag {\tt luvcheck}). The same check for the
infrared (IR) behaviour ({\tt lricheck}) will change the explicit
values in columns {\tt loop} and {\tt soft}, but the sum should still
be invariant.
 The last column may also suffer against IR variation
since it is neither compensated by double soft photon emission nor
two-loop divergencies. Those {\verb+loop^2+} numbers are crucial when they become the leading order
in e.g.\ flavour changing neutral current processes (FCNC) as
discussed in \cite{Lorca:2004dk}, or just as an estimator of the order of magnitude for the error in the next level of perturbation theory.
When integrated cross sections are under study, the integration region
is limited by the {\tt setlimcost} variable, the {\tt ics} and {\tt
  fba} being defined in terms of {\tt dcs} as
\bea
\mathtt{dcs}&=&\frac{\d{\sigma}}{\d\cos{\theta}}\\
\mathtt{ics}&=&\sigma_\mathrm{tot}=
\int_{-\mathtt{setlimcost}}^{\mathtt{setlimcost}}
\!\!\!\!\d\cos{\theta}~ \mathtt{dcs}
\label{ics}\\
\mathtt{fba}&=&\frac{\sigma_\mathrm{fw} - \sigma_\mathrm{bw}}{\sigma_\mathrm{tot}}=
\int_{0}^{\mathtt{setlimcost}} \!\!\!\! \d\cos{\theta}~ \frac{\mathtt{dcs}}{\mathtt{ics}}  -
\int_{-\mathtt{setlimcost}}^{0} \!\!\!\! \d\cos{\theta}~
\frac{\mathtt{dcs}}{\mathtt{ics}}.
\label{dcs}
\eea

The integration algorithm is based on a Richardson extrapolation to
the Romberg integration \cite{Richardson} with four steps. The estimated error is supplied in
short forms in extra columns. If it does not suffice or for cross
checking, the same subroutine with eight steps ({\tt ICS8}) can be
called instead of the faster {\tt ICS}.
A complete survey of variables is presented in \tab{mainsettings}.

\subsection{A greedy possibility: quadruple precision}
Not every compiler\footnote{Until now, the use of quadruple precision
  has been positively tested under {\sc Intel} \fortran\ compiler (see
\\ \url{http://www.intel.com/software/products/compilers/flin/noncom.htm}).}
 allows for extended or quadruple precision. This is a feature
outside of the ANSI \fortran 77. Even if \aitalc\ contains some
code's implementations outside the ANSI syntax, they are quite common
(e.g. using underscore, or double complex), so decent compilers won't complain.

But on high level computation, or simply as comparison, switching on the quadruple precision could stabilize numerical results or render amazing agreement with other calculations.


\section{Examples including loop level}

To finish this descriptive chapter, we would like to present the examples available with the distribution.

The package is delivered with three examples that the user may run without further considerations. In the directory {\tt examples/} we have the following processes:
\bi
\item \textmu-pair production at tree level: $e^- e^+ \rightarrow \mu^- \mu^+$
\item Bhabha scattering in QED (see \cite{Gluza:2004tq}): $e^- e^+ \rightarrow e^- e^+$
\item Fermion Flavour Violation example:  $e^- e^+ \rightarrow b \bar{s}$
\ei
under the directory names of {\tt muon\_production}, {\tt Bhabha\_QED} and {\tt leLe.bS}.

\begin{table}[h]
\begin{center}
\caption{Typical timings for the different modules. The technical specifications for system Laptop are: Intel Centrino 1.5GHz cpu, 512MB RAM, Intel {\tt icc} and {\tt ifort} version 8.1 compilers. For Desktop: Intel Pentium III 853MHz cpu, 256MB RAM, GNU {\tt gcc} and {\tt g77} version 3.3.3 compilers.}
\label{timings}
\begin{tabular}{|l|rr|rr|rr|}
\hline
&\multicolumn{2}{|c|}{\tt
  muon\_production}&\multicolumn{2}{|c|}{\tt
  leLe.bS}&\multicolumn{2}{|c|}{\tt bhabha\_QED}\\
Module&Desktop&Laptop&Desktop&Laptop&Desktop&Laptop\\
\hline
\hline
{\tt tree}&9''&3''&4''&1''&30''&9''\\
{\tt loop}&--\phantom{''}&--\phantom{''}&2'20''&52''&37''&12''\\
{\tt fortran}&15''&4''&4:35'56''  &33''&26''&8''\\
\hline
Total&24''&7''&4:38'20''&1'26''&1'33''&29''\\
\hline
\end{tabular}
\end{center}
\end{table}

Running time expected for the execution of all the examples were studied on two
different computers. The results are shown in \tab{timings}. One can see
that when the size of the \fortran\ code begins to be large (i.e.\ more than $\sim$200kB for a single subroutine to be compiled), then
the required time for the GNU compiler gets too long, compared to the
other processes or compiler.


 \appendix

\section{Implemented models}
\begin{table}[h]
\caption{Particle content of the QED model ({\tt QED.model})}
\label{qedtable}
\begin{center}
\begin{tabular}{|c@{, }ccccc|}
\hline
\multicolumn{2}{|c}{Code}&Arguments&Particle&Name&Lorentz type\\
\hline
\hline
\multicolumn{6}{|l|}{{\sl Lepton}}\\
{\tt le}&{\tt Le}&{\tt(;$p_i$)}&$e^-$, $e^+$&electron, positron&fermion\\
\hline
\multicolumn{6}{|l|}{{\sl Gauge boson}}\\
\multicolumn{2}{|c}{\tt A}&{\tt($\mu_i$;$p_i$)}&$\gamma$&photon&boson\\
\hline
\end{tabular}
\end{center}
\end{table}
\newpage

\begin{table}[ht]
\caption{Particle content of the electro-weak sector in the standard
  model ({\tt EWSM.model})}
\label{ewsmtable}
\begin{center}
\begin{tabular}{|c@{, }ccccc|}
\hline
\multicolumn{2}{|c}{Code}&Arguments&Particle&Name&Lorentz type\\
\hline
\hline
\multicolumn{6}{|l|}{{\sl Leptons}}\\
{\tt ne}&{\tt Ne}&&$\nu_e$, $\bar{\nu}_e$&neutrino-$e$, antineut-$e$&\\
{\tt nm}&{\tt Nm}&{\tt(;$p_i$)}&$\nu_\mu$, $\bar{\nu}_\mu$&neutrino-$\mu$, antineut-$\mu$&fermion\\
{\tt nt}&{\tt Nt}&&$\nu_\tau$, $\bar{\nu}_\tau$&neutrino-$\tau$, antineut-$\tau$&\\
{\tt le}&{\tt Le}&&$e^-$, $e^+$&electron, positron&\\
{\tt lm}&{\tt Lt}&{\tt(;$p_i$)}&$\mu^-$, $\mu^+$&muon, antimuon&fermion\\
{\tt lt}&{\tt Lt}&&$\tau^-$, $\tau^+$&tau, antitau&\\
\hline
\multicolumn{6}{|l|}{{\sl Quarks}}\\
{\tt u}&{\tt U}&&$u$, $\bar{u}$&up, antiup&\\
{\tt c}&{\tt C}&{\tt(;$p_i$)}&$c$, $\bar{c}$&charm, anticharm&fermion\\
{\tt t}&{\tt T}&&$t$, $\bar{t}$&top, antitop&\\
{\tt d}&{\tt D}&&$d$, $\bar{d}$&down, antidown&\\
{\tt s}&{\tt S}&{\tt(;$p_i$)}&$s$, $\bar{s}$&strange, antistrange&fermion\\
{\tt b}&{\tt B}&&$b$, $\bar{b}$&bottom, antibottom&\\
\hline
\multicolumn{6}{|l|}{{\sl Gauge bosons}}\\
\multicolumn{2}{|c}{\tt A}&&$\gamma$&photon&\\
\multicolumn{2}{|c}{\tt Z}&{\tt($\mu_i$;$p_i$)}&$Z$&Z-boson&vector\\
{\tt Wm}&{\tt Wp}&&$W^-$, $W^+$&W-boson&\\
\hline
\multicolumn{6}{|l|}{{\sl Higgs sector}}\\
\multicolumn{2}{|c}{\tt H}&&$H$&Higgs&\\
\multicolumn{2}{|c}{\tt G0}&{\tt(;$p_i$)}&$\chi$&would-be&scalar\\
{\tt Gm}&{\tt Gp}&&$\phi^-$, $\phi^+$&Goldstone bosons&\\
\hline
\multicolumn{6}{|l|}{{\sl Faddeev-Popov ghosts}}\\
{\tt ghA}&{\tt GhA}&&$\eta_\gamma$, $\bar{\eta}_\gamma$&photon ghosts&scalar\\
{\tt ghZ}&{\tt GhZ}&{\tt(;$p_i$)}&$\eta_Z  $, $\bar{\eta}_Z  $&Z-ghosts&with\\
{\tt ghm}&{\tt Ghm}&&$\eta_{W^-}$,$\bar{\eta}_{W^-}$&W$^-$-ghosts&fermion\\
{\tt ghp}&{\tt Ghp}&&$\eta_{W^+}$, $\bar{\eta}_{W^+}$&W$^+$-ghosts&statistics\\
\hline
\end{tabular}
\end{center}
\end{table}
\newpage

\section{List of \protect \form{} procedures}
\begin{center}
\tablefirsthead{%
\hline
Name & Args. & Description \\
\hline
\hline
}
\tablehead{%
\hline
\multicolumn{3}{|l|}{\small \sl continued from previous page}\\
\hline
Name & Args. & Description \\
\hline
\hline
}
\tabletail{
\hline
\multicolumn{3}{|l|}{\small \sl continued on next page}\\
\hline
}
\tablelasttail{\hline}
\tablecaption{Different procedures used.}
\label{procedurelist}
\begin{supertabular}{|lcp{18em}|}
{\tt LTtoPV }& & Translates the indexing of loop integrals from \cite{Denner:1991kt}
into \cite{'tHooft:1978xw} convention.\\
{\tt aitalcnotation }& $N$ & Translates the \form{} code into
\fortran{} readable. The argument $N=1$ allows for complex masses
while $N=0$ does not\\
{\tt analyzeterms }& & Part of {\tt simplifygrams} \\
{\tt argsymmetries }& & Applies symmetry properties on arguments for
loop functions and Gram determinants\\
{\tt canceldens }& & To cancel composition of {\tt den(a,b)*(a-b)=1}\\
{\tt chisholm }& & Applies Chisholm identities by converting all chains
of $\gamma$ matrices that are contracted by their dimensional value \\
{\tt contgammas }& & Simplifies contractions in products of gamma matrices\\
{\tt contractepsilon }& & It contracts the product of two $\epsilon$ tensors\\
{\tt ctfeynmanrules }& & Insertion of Feynman rules with counterterms\\
{\tt definegrams }& & Defines a \form{} global variable with the explicit
expression of a Gram determinant\\
{\tt derivateself }& & Derivates self-energy functions\\
{\tt dimensionfour }& & To reach the limit $D \to 4$ taking care of the
UV-behaviour of scalar integrals\\
{\tt diracequation }& & Applies Dirac equation to available spinors\\
{\tt dummytovar }& $N$ & Part of {\tt simplifygrams}. $N$ stands for
the amount of arguments in {\tt dummyN} function \\
{\tt equalindex }& & Gives the same index to potentially different contractions keeping them ordered by number \\
{\tt externalmomenta }& & To substitute the internal momenta in loops with external ones\\
{\tt gamma3to1 }& & Introduces an $\epsilon$ tensor and $\gf$ to
remove the product of three $\gamma$'s\\
{\tt gammaalgebra }& & Prepares $\gamma$ chain structures for {\tt diracequation}\\
{\tt gammamunu }& & Introduces indices in the vector structures\\
{\tt gramback }& $N$ & Part of {\tt simplifygrams}. Returns the Gram
determinant of order $N$ if no simplification
was found\\
{\tt gramsubstitution }& $N,w$ & It substitutes the $w$-element to
match posssible order $N$ Gram determinants\\
{\tt identifyintegrals }& & Picks up the different loop integrals and
provides a set of \mbox{ \form{} {\tt \$}-variables} with the right substitutions,
inserting the function {\tt LoopIntegral=LI(n)} \\
{\tt integrationde93 }& & Procedure to integrate 4,3,2,1-point
functions with internal momenta decomposition like in \cite{Denner:1991kt}\\
{\tt invgammamunu }& & Undoes the {\tt gammamunu} effects\\
{\tt keepUVloops }& & Keeps only integrals actually UV divergent\\
{\tt loopsymmetries }& & Minimize loop integrals in boxes by
symmetrizing arguments\\
{\tt lorentzinvariants }& & Adapts scalar product of vectors to Lorentz invariant squared masses\\
{\tt massconvention }& & To use single variables for the masses with {\tt M} instead
of {\tt MM} outside arguments\\
{\tt massivefofa }& $j,c,N$ & Calculates and saves the formfactors as
\cite{Lorca:2005}. {\tt j} stands for topology, {\tt c} for left-right or $\gi$-$\gf$ and {\tt N=1(0)} do (not) print
the formfactors \\
{\tt movepslash }& & Places the desired $\slashed{p}_i$ to the side of
fermion chain to apply {\tt diracequation}\\
{\tt neglectmass }& {\tt zero} & Neglects positive powers of terms as indicated in the
file {\tt Neglectedmasses.inc}. {\tt zero} is the substituted value \\
{\tt nosymboliccouplings }& & To put explicit values of charges and weak couplings \\
{\tt onshell }& & Substitutes external momenta for external particle masses and Mandelstam variables \\
{\tt pslashaway }& & Reorder away the non desired position of $\slashed{p}_i$\\
{\tt pushgamma5 }& & Pushes $\gf$ to the right in fermion chains \\
{\tt pushomegas }& & Orders and pushes Left or Right projectors ($\omega_{L,R}$) to the right in fermion chains\\
{\tt recoverargumentsde93 }& & Catchs back the arguments of loop
integrals in combination with {\tt integrationde93}\\
{\tt reductionDD02 }& $N$ &  Reduction algorithm from scalar tensor
integrals to master integrals given in \cite{Denner:2002ii}. $N=1$
sorts after each integral type, $N=0$ does not sort at all\\
{\tt reductionLT }& & {\tt reductionDD02(1)} with \LT{} notation  \\
{\tt simplifygrams }& & Tries to remove inverse Gram determinants by
looking to the numerator (time and memory consuming!)\\
{\tt simplihelp }& $N,r$ & It helps {\tt simplifygrams} to run
sequentially $r$ times from the highest power of one variable for $N$
order Gram determinants\\
{\tt storeself }& $j,c,N$ & Stores self energies, {\tt j} stands for topology, {\tt c} for
($\omega_{L}$,$\omega_{R}$) or ($\gi$,$\gf$) and {\tt N=1(0)} do (not) print
the form factors\\
{\tt threegammastoepsilon }& & Introduces and removes the $\epsilon$ tensor simplifying chains of 3 $\gamma$ matrices that naturally don't dissappear after {\tt diracequation}\\
{\tt tracefermiloops }& & Traces possible fermion loops in self-energies\\
{\tt transvorlongit }& $x$ & Sets the longitudinal and transverse part
of self energies, keeping off-shell but with mass$^2$ argument $x$\\
{\tt unityCKM }& & Makes $\mathrm{CKM}_{ij}= \delta_{ij}$ \\
{\tt usegamma5 }& & Uses $\gf$, instead of $\omega_{L,R}$ projectors\\
{\tt useomegas }& & The opposite to {\tt usegamma5}\\
{\tt vartodummy }& & Part of {\tt gramback}, opposite as {\tt dummytovar}\\
\end{supertabular}
\end{center}
\newpage

\section{Settings for {\tt main.f}}
\def \INT{integer}
\def \DP{double}
\def \LOG{logical}

\begin{center}
\tablefirsthead{%
\hline
Variable&Type&Def.\ value&Meaning\\
\hline
\hline
}
\tablehead{%
\hline
\multicolumn{4}{|l|}{\small \sl continued from previous page}\\
\hline
Variable&Type&Def.\ value&Meaning\\
\hline
\hline
}
\tabletail{
\hline
\multicolumn{4}{|l|}{\small \sl continued on next page}\\
\hline
}
\tablelasttail{\hline}
\topcaption{Kinematical and variables and logical flags for different settings in {\tt main.f}.}
\label{mainsettings}
\begin{supertabular}{|llcp{14.5em}|}
\hline
\multicolumn{4}{|l|}{\sl {Kinematical}}\\
{\tt icostloop}&\INT&{\tt 5}&Dimension of {\tt setcostarray}\\
{\tt isqrtsloop}&\INT&{\tt 5}&Dimension of {\tt setsqrtsmanarray}\\
{\tt setcostarray}&\DP&{\tt data //}&Set of default values for the
{\tt setcost} ($\cos{\theta}$) do loop in the {\tt dcs} evaluation\\
{\tt setsqrtsmanarray}&\DP&{\tt data //}&Set of default values for the
{\tt setsqrtsman} ($\sqrt{s}$) do loop in the {\tt ICS} evaluation\\
{\tt setfracomega} &\DP &{\tt 0.1d0}& Limit for the maximum soft
photon energy\\
{\tt setlimcost}&\DP &{\tt .9999d0}& Limit of integration for {\tt
  ics} and {\tt fba}\\
\hline
\multicolumn{4}{|l|}{\sl {Process flags}}\\
{\tt lrenorm}&\LOG&{\tt .true.}& Call the renormalization subroutine
to calculate the counterterms parameters\\
{\tt lwidth}&\LOG&{\tt .false.}& Call the loop integrals with complex
values for boson masses. For the four-point integral, only the case
with one massless boson (photon) is considered.\\
{\tt lidentCKM}&\LOG&{\tt .false.}& Brings the CKM mixing matrix in a
diagonal form, so no mixing occurs\\
\hline
\multicolumn{4}{|l|}{\sl {Checking flags}}\\
{\tt luvcheck}&\LOG&{\tt .false.}& Perform a shift in the {\tt mudim}
and {\tt delta} parameters in \LT\\
{\tt lircheck}&\LOG &{\tt .false.}& Perform a shift in the {\tt
  lambda} parameter in \LT\\
\hline
\multicolumn{4}{|l|}{\sl {Output flags}}\\
{\tt lcostloop}&\LOG &{\tt .true.}& Performs a do-loop over {\tt setcost}\\
{\tt lprintics}&\LOG&{\tt .false.}& Performs a do-loop over {\tt
  setsqrtsman} and prints the results of {\tt ics}\\
{\tt lprintfba}&\LOG&{\tt .false.}& Performs a do-loop over {\tt
  setsqrtsman} and prints the results of {\tt fba}\\
{\tt llongoutput}& \LOG&{\tt .false.}& Prints the cross sections in
double precision format\\
\end{supertabular}
\end{center}

\end{document}